# Genomic 3D-compartments emerge from unfolding mitotic chromosomes

Rajendra Kumar[1,2], Ludvig Lizana[1,2]*, Per Stenberg[3,4]*

[1] Integrated Science Lab, Umeå University, Sweden.

[2] Department of Physics, Umeå University, Sweden.

[3] Department of Ecology and Environmental Science (EMG), Umeå University, Sweden.

[4] Division of CBRN Security and Defence, FOI-Swedish Defence Research Agency, Umeå, Sweden

* To whom correspondence should be addressed. Email: ludvig.lizana@umu.se, per.stenberg@umu.se (+46 90 7856777)

Acknowledgements

This project was funded by the Kempe Foundation (grant number JCK-1347 to LL and PS) and the Knut and Alice Wallenberg foundation (grant number 2014-0018, to EpiCoN, co-PI: PS). The simulations were performed on resources provided by the Swedish National Infrastructure for Computing (SNIC) at HPC2N Umeå, Sweden.

Abstract

The 3D organisation of the genome in interphase cells is not a randomly folded polymer. Rather, experiments show that chromosomes arrange into a network of 3D compartments that correlate with biological processes, such as transcription, chromatin modifications, and protein binding. However, these compartments do not exist during cell division when the DNA is condensed, and it is unclear how and when they emerge. In this paper, we focus on the early stages after cell-division as the chromosomes start to decondense. We use a simple polymer model to understand the types of 3D structures that emerge from local unfolding of a compact initial state. From simulations, we recover 3D compartments, such as TADs and A/B compartments, that are consistently detected in Chromosome Capture Experiments across cell types and organisms. This suggests that the large-scale 3D organisation is a result of an inflation process.

Introduction

Apart from the bare challenge of packing a long DNA polymer into a small cell nucleus without heavy knotting, the DNA must fold in 3D to allow nuclear processes, such as gene activation, repression and transcription, to run smoothly. By how much the DNA folding patterns influences these processes, and by how much they influence human health, is currently attracting a lot of attention in the scientific community (Cremer and Cremer 2001; Fullwood et al. 2009; Gondor 2013; Krijger and de Laat 2016; Schneider and Grosschedl 2007; Sexton et al. 2007).

To better understand DNA's 3D organization, researchers developed various Chromosome Conformation Capture Methods. The most recent incarnation, the Hi-C method (Lieberman-Aiden et al. 2009), measures contact probabilities between all pairs of loci in the genome. Across cell types and organisms, Hi-C repeatedly detects two types of co-existing megabase scale structures. First, all chromosome loci seem to belong to one of two so-called A/B compartments (Lieberman-Aiden et al. 2009), where the chromatin in one compartment is generally more open, accessible, and actively transcribed than the other. Second, linear subsections of the genome assemble into topological domains (Dixon et al. 2012; Nora et al. 2012), often referred to as Topologically Associating Domains (TADs), that show up in the Hi-C data as local regions with sharp borders with more internal than external contacts. These borders correlate with several genetic processes, such as transcription, localization of some epigenetic marks, and binding positions of several proteins – most notably CTCF and Cohesin (Dixon et al. 2012; Nora et al. 2012)). However, even though researchers established these correlations, we still lack a general mechanistic understanding for how TADs and A/B compartments form.

To figure out these mechanisms experimentally poses a big challenge. Several research groups have therefore turned to computer models (Dekker et al. 2013; Rosa and Zimmer 2014). Apart from so-called restraint based models that optimise 3D distances between all DNA fragments using Hi-C data (Fraser et al. 2009), theorists often represent DNA as a polymer fibre (Barbieri et al. 2012; Mirny 2011; Sachs et al. 1995; Therizols et al. 2010). One example is the fractal globule (Grosberg et al. 1993), a compact and knot-free polymer, that is compatible with looping probabilities in the first human Hi-C experiment (Lieberman-Aiden et al. 2009). However, recent work (Sanborn et al. 2015) cast doubt on some of the model's predictions because (1) the looping probability exponent varies on small and large scales (as well as during the cell cycle) and (2) it cannot be used to understand TADs or A/B compartments because fractal globules lack domains. To bridge this gap, researchers developed several mechanistic models. For example, Sanborn et al. (2015) used a ring-like protein (Cohesin) that pulls the DNA trough itself until it reaches a CTCF-site where it stops. In another example (Barbieri et al. 2012; Fraser et al. 2015), the authors used a polymer with binding sites to particles that diffused in the surrounding volume. As these particles may simultaneously bind to several sites, they stabilise loops and create nested TADs.

However, while these models can predict TAD-like structures that are formed by loop-stabilizing protein complexes, such as CTCF and Cohesin, they do not explain A/B compartments. And furthermore, it is unclear if all TADs are loops at all. Moreover, most polymer and restraint-based approaches initially prepare the system in some random configuration and let it equilibrate. With the right set of conditions, the system then folds into domains such as TADs. But this is far from how the process happens in the cell. Just after cell division, the chromosomes are about 4-50 times more compact on the linear scale (where chromatin that is more open during interphase shows the highest difference) and occupy roughly half the volume than when unfolded during interphase (Belmont 2006; Li et al. 1998; Mora-Bermudez et al. 2007). In addition, mitotic chromosomes seem to lack any clear domain structure (Naumova et al. 2013). This suggests that all domains emerge as chromosomes unfolds. This aspect is overlooked in most models. To better understand the types of structural compartments that can emerge from a compact initial state, we used simulations to study the unfolding process of a polymer as sub-sections decondense. We find that both TADs and A/B compartments can form without the need to introduce loop-stabilising attractors.

Results and Discussion

We model a chromosome as a beads-on-a-string polymer where each bead represents a piece of chromatin. Apart from nearest neighbour harmonic bonds (i.e. hookean springs), the beads attract each other via a Lennard-Jones potential that also prevents the beads from overlapping. To construct a compact polymer that mimics a mitotic chromosome, we used the GROMACS molecular dynamics package to crumple the polymer into a globule under the Lennard-Jones potential (Fig. 1a). Similar to real Hi-C data on mitotic chromosomes (Naumova et al. 2013) our simulated globule lacks domain structure (see Supporting Fig. S1).

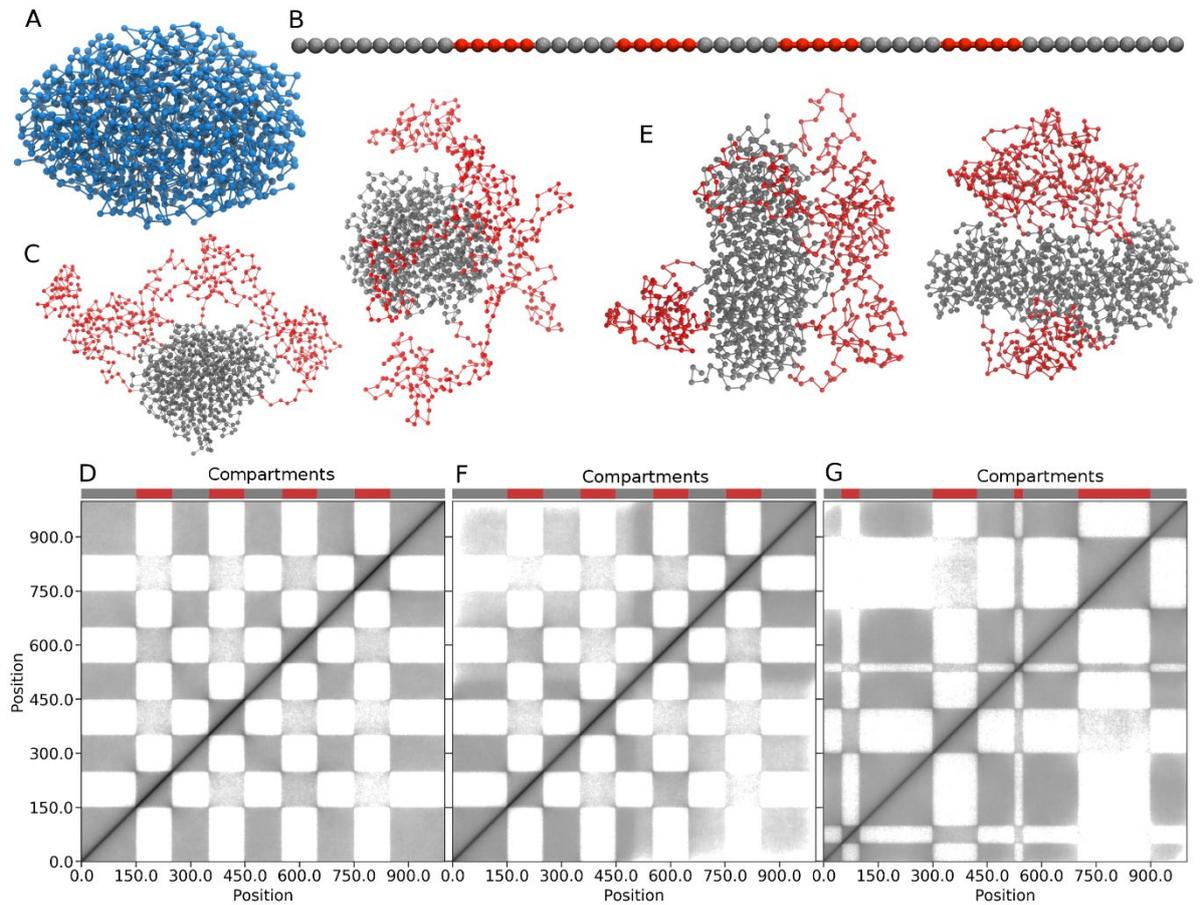

**Figure 1. 3D domains emerge from local unfolding of a compact polymer.** (A) An example of a simulated compact polymer. (B) Schematic representation of open (red) and compact (grey) regions (in the simulations we used 1,000 beads). (C) Two examples of unfolded polymers starting from a spherical initial condition (no enforced globule elongation). (D) Average bead-bead contact map obtained from an ensemble of polymer structures as those in (A). Note the checkerboard pattern. (E) Two unfolded polymers from a cigar-shaped mitotic chromosome-like initial condition (with enforced globule elongation) (F) Average bead-bead contact map obtained from an ensemble of polymer structures as those in (E). Note the intensity decay with increasing distance from the diagonal. (G) Contact map where open and compact regions have different lengths.

To model the unfolding from the crumpled state, as for example when genes turn on, we partitioned the crumpled polymer into two types of regions that alternate along the polymer (Fig. 1b). Labelled as red and grey, the red parts are more flexible than the grey ones. In our simulations, we achieve this by lowering the Lennard-Jones interaction potential $V(r)$ between red beads (separated by the distance *r*). In more detail, we lowered the energy scale ε in $V(r) = 4\varepsilon\left[\left(\frac{\sigma}{r}\right)^{12} - \left(\frac{\sigma}{r}\right)^{6}\right]$, to represent a lower "stickiness". For example, compact heterochromatin is considered stickier compared to open chromatin. However, the exact reasons behind this is not completely understood but some studies indicate that histone modifications and HP1 is involved (Antonin and Neumann 2016; Hug et al. 2017; Maison and Almouzni 2004). Finally, the parameter σ is the distance where *V(r=σ)* is zero.

To determine the relative values of ε for different chromatin types, we calculated the radius of gyration as a measure of compactness for polymers where all beads were of the same type (Supporting Fig. S2). During crumpling, we use ε=2.5 to achieve a condensed globule (Fig. 1A). During the decondensation stage, to reduce computational time when generating a large number of diverse crumpled configurations, we lowered ε to 1.5. This is the highest value of epsilon before the globule starts to unfold (Supporting Fig. S2). This means that ε must be lower than 1.5 for the open chromatin state. We choose ε=0.75 for two reasons: (1) If ε is close to 1.5, there will be very little decompaction. (2), If ε is too small the volume that the unfolded polymer occupies will quickly be very large. In fact, we found that at ε=0.75 the volume change from the crumpled globule to the decondensed state was roughly two-fold (Supporting Fig. S3), which is similar to experimental observations (Mora-Bermudez et al. 2007). However, it should be noted that our simulation does not include a volume barrier, which for real chromosomes would be the nuclear envelope. In Table 1, we summarize the parameter values we used in $V(r)$ during different stages of our simulations.

**Table 1: Lennard-Jones parameters used for condensation and decondensation (GROMACS' default unit).**

| Bead-pair type | σ | ε |
|---|---|---|
| Condensation - linear chain to globule | | |
| bead-bead | 0.178 | 2.5 |
| Decondensation - unfolding of globule | | |
| close-close | 0.178 | 1.5 |
| open-open | 0.178 | 0.75 |
| open-close | 0.178 | 0.05 |

After crumpling and partitioning, we simulated how the polymer unfolds under thermal fluctuations. Figure 1c shows two snapshots of a simulated polymer. As in all realisations we investigated, these show that the red flexible parts are on the exterior of the polymer, whereas the grey parts remain compact. We stop the simulation after 1,000,000 MD-steps where the red parts are clearly decondensed, and then store the structure for analysis. To rapidly generate diverse polymer configurations, we used periodic simulated annealing (see Fig. 2 and details in Supporting Fig. S4).

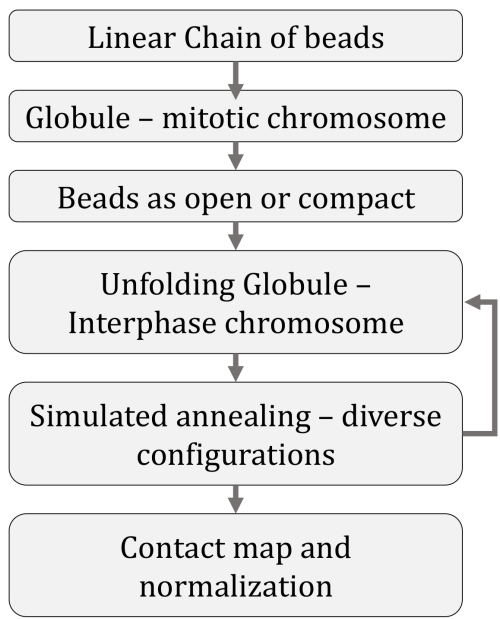

**Figure 2. Summary of the work-flow for heterogenous unfolding of the compact polymer.** A more detailed flow-chart is provided in Supporting Fig. S1.

With the unfolding mechanism in place, we generated an ensemble of unfolded polymers (1,000 beads each), all starting from different realizations of the compact globule (Supporting Fig. S4), and then measured the distance between all bead pairs. If the distance between beads' centres was shorter than two times the beads' diameter, we defined it as a physical contact. Collecting all contacts, we made an artificial Hi-C map and normalized it with the KR-norm (Knight and Ruiz 2013), as in real Hi-C experiments. Finally, we visualised the artificial Hi-C map in the gcMapExplorer software (Kumar et al. 2017) (Fig. 1d).

Two things stand out when looking at Fig. 1d: (i) the TAD-like structure along the diagonal, (ii) the off diagonal plaid pattern that resembles A/B compartments. These two are universal features of all experimental Hi-C maps and appears also here. We get these patterns from a minimal set of assumptions. In particular, without specific chromatin binding proteins.

However, we observe that the contact frequency in Fig. 1d does not decay as a function of the linear distance between beads (the off-diagonal direction). Apart from short distances, this is not consistent with real Hi-C maps where the intensity decays roughly as a power- law with distance (Lieberman-Aiden et al. 2009). The reason is that we used a simple simulation protocol that produces spherically shaped starting configurations (Fig. 1d). To remedy this, we added a global potential (see methods) that gives a cigar-like globule (Fig. 1e). Notably, we do not argue that this is how the mitotic chromosome get its shape in the cell. It is a pragmatic way to get a starting configuration that is more realistic than a sphere. With this modification to the simulation protocol we get an intensity that decays with linear distance between bead pairs (Fig. 1f). To further make our system more realistic, we acknowledge that open and compact regions along chromosomes do not have the same length. By varying the length of these in the simulations, the plaid patterns in the contact map (Fig. 1g) approach even more those we observe in real Hi-C maps.

To conclude, we show that partial de-condensation of a simple mitotic-chromosome-like polymer is enough to recreate TADs, A/B compartments, and contact frequency decay over distance - universal features of all (interphase) Hi-C maps across cell types and organisms. Although, our results do not exclude that specific loop-forming proteins are essential to shape and maintain the genomes' 3D structure, our work underscores that chromosomes' large scale 3D organisation is the result of an inflation process. We look forward to the next-generation 3D genome models that integrate specific interactions, such as loop-stabilising protein complexes and chromatin states, with the initial compact chromosome state.

Methods

We simulated a linear polymer in the GROMACS molecular dynamics package where the beads (or monomers) interact via a Lennard-Jones potential (for convenience we set the bead radius to 1 Å to reduce the problem to atomic scales): $V(r) = 4\varepsilon\left[\left(\frac{\sigma}{r}\right)^{12} - \left(\frac{\sigma}{r}\right)^{6}\right]$. To form a globule, the value of ε was set so that that the resulting attractive force between beads would overcome thermal fluctuations. During decondensation, the value of ε was reduced by 40%, 70% and 98% for interaction between close-close, open-open and close-open beads, respectively. The value of σ was kept constant throughout the simulation (see Table 1).

To condense the polymer into a compact globule, we used GROMACS' Langevin dynamics module. Since creating a large globule (1000 beads) takes time, we made two 500 bead globules and mixed those (Supporting Fig. S4). We then used several cycles of simulated annealing (Supporting Fig. S4) to obtain diverse globule configurations. Furthermore, since the polymers' ends are free, there could be problems with reptation and subsequent knot formation (the mitotic chromosome is largely unknotted). To prevent this, the globules' ends were capped by a 10 beads-on-string terminal containing a stiff angular harmonic restraint to prevent bending at the two terminals. To efficiently explore as much of the conformational space as possible, we used a periodic simulated annealing approach detailed in Supporting Fig. S4. The GROMACS parameters we used for the simulations are listed in Table 2.

**Table 2: GROMACS MD parameters used during different stages.** All parameter values were kept constant, except for the MD steps, which were different depending on the stages.

| Parameter | Value |
|---|---|
| integrator | bd |
| dt | 0.001 ps |
| steps | 1000000000 |
| **Langevin Dynamics Options** | |
| bd-fric | 0 |
| ld-seed | -1 |
| **Neighbour searching Parameters** | |
| cutoff-scheme | group |
| nstlist | 1 |
| rlist | 2 |

| Options for van der Waals | |
|---|---|
| vdw-type | Cut-off |
| rvdw | 2 nm |
| **Temperature coupling** | |
| Tcoupl | v-rescale |
| nsttcouple | 1 |
| tau_t | 0.001 ps |
| ref_t | 200K |

The above simulation protocol leads to a spherically shaped object. However, the mitotic chromosome is elongated rather than spherical. To achieve this, we used so-called steered MD simulation with zero pulling velocity. Simply put, we introduced a harmonic pull potential between the centre of masses between the two 500 bead globules while they mixed. After globule formation, we let the globule unfold under thermal fluctuations. In the flexible regions (red beads) we lower the Lennard-Jones parameters compared to the compact region (grey). We show these and all other GROMACS force-field parameters in Table 1 and Table 2.

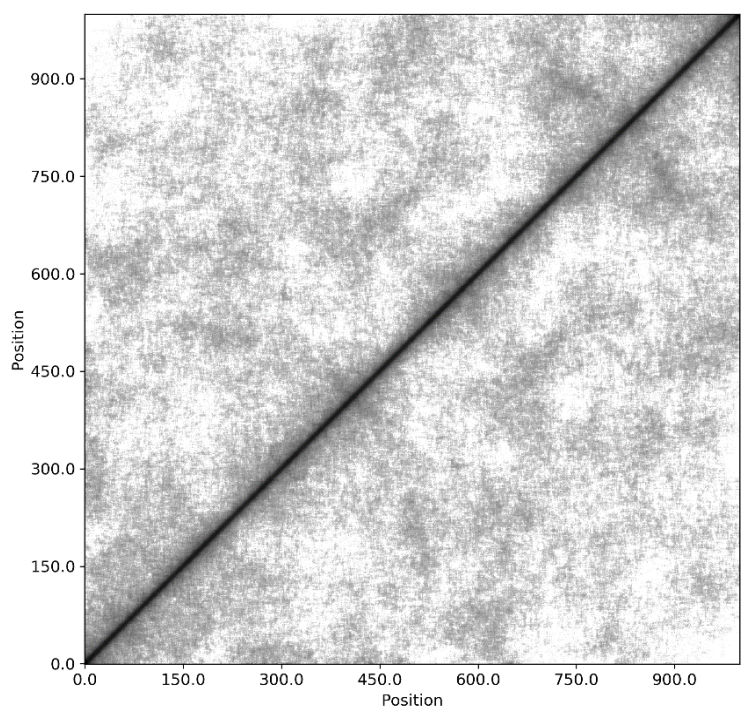

**Figure S1: Contact map of crumpled globules**. Average bead-bead contact map obtained from an ensemble of crumpled globules.

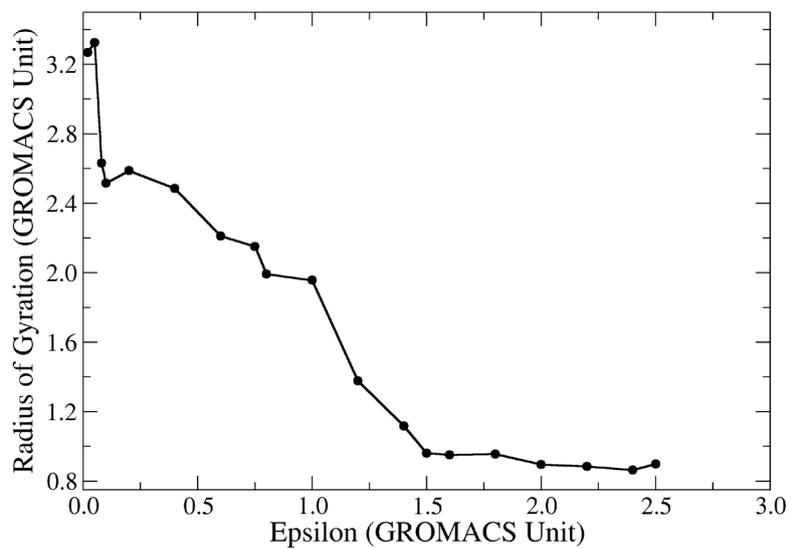

**Figure S2: The radius of gyration as a function of epsilon**. The black line is based on an average of 10 globule configurations at each epsilon value. In this simulation all beads in the polymer are identical.

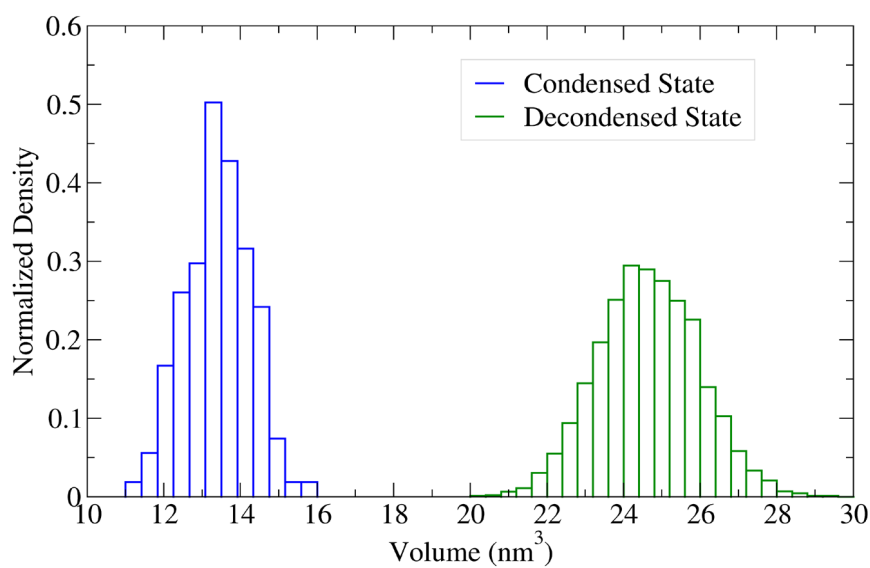

**Figure S3: Volume of the condensed and decondensed state**. Volume distribution for realized configurations of globules at condensed and decondensed states were calculated from the simulations. To calculate volume of the globules, we considered a probe of 2 Å to create a probe accessible surface enclosing the globule and calculated volume of this enclosed shape very similar to solvent accessible surface for macromolecules using the GROMACS tool g_sas. The probe of 2 Å radius was considered because 4 Å was used to calculate the contact map and with this radius, two beads in contact will be inside the same enclosed surface when rolling the probe over the globule.

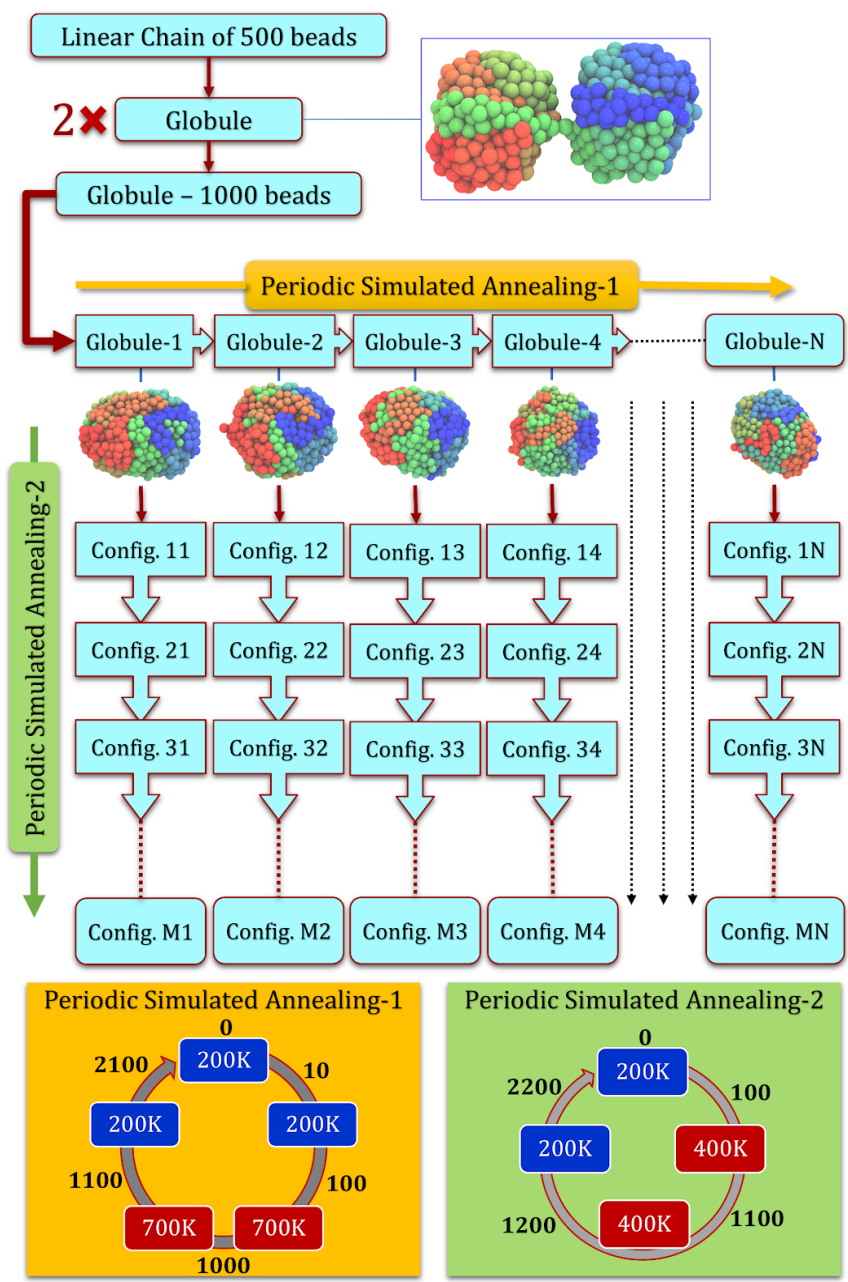

**Figure S4: A schematic illustration of the simulation procedure.** First we form two 500 bead globules, connect them, and merge them. The monomers are coloured from read through green to blue according to their linear order in the polymer. After merger, we used periodic simulated annealing (highlighted in orange) to generate diverse globule configurations. Subsequently, each globule was unfolded and periodic simulated annealing (green) was performed to generate diverse polymer configurations. At bottom we show the annealing process. The numbers inside the boxes show effective temperatures (Kelvin) while arrow indicates the number of MD steps.